\documentclass[twocolumn,superscriptaddress,nobibnotes,footinbib,amsmath,floatfix,aps,prapplied,citeautoscript,reprint]{revtex4-1}

\usepackage{graphicx}
\usepackage{dcolumn}
\usepackage{bm}
\usepackage{amsmath,bm,multirow}
\usepackage{amsfonts}
\usepackage{float}
\usepackage{color}
\usepackage[draft]{changes} 
\usepackage[normalem]{ulem}

\newcommand{\psu}{Department of Mechanical and Materials Engineering, Portland State University, Portland, OR 97201, USA}

\newcommand{\nw}{Department of Materials Science and Engineering, Northwestern University, Evanston, IL 60208, USA}

\usepackage{float}
\usepackage{graphicx}
\usepackage{sidecap}
\usepackage{tabularx}
\usepackage{sidecap}
\newcommand{\etal}{\textit{et al}. }

\begin{document}
\title{Accelerating High-Throughput Phonon Calculations via Machine Learning Universal Potentials}

\author{Huiju Lee}
\affiliation{\psu}

\author{Vinay I. Hegde}
\affiliation{\nw}

\author{Chris Wolverton}
\affiliation{\nw}

\author{Yi Xia}
\email{yxia@pdx.edu}
\email{yimaverickxia@gmail.com}
\affiliation{\psu}

\date{\today}

\begin{abstract}
Phonons play a critical role in determining various material properties, but conventional methods for phonon calculations are computationally intensive, limiting their broad applicability. In this study, we present an approach to accelerate high-throughput harmonic phonon calculations using machine learning universal potentials. We train a state-of-the-art machine learning interatomic potential, based on multi-atomic cluster expansion (MACE), on a comprehensive dataset of 2,738 crystal structures with 77 elements, totaling 15,670 supercell structures, computed using high-fidelity density functional theory (DFT) calculations. Our approach significantly reduces the number of required supercells for phonon calculations while maintaining high accuracy in predicting harmonic phonon properties across diverse materials. The trained model is validated against phonon calculations for a held-out subset of 384 materials, achieving a mean absolute error (MAE) of 0.18~THz for vibrational frequencies from full phonon dispersions, 2.19~meV/atom for Helmholtz vibrational free energies at 300K, as well as a classification accuracy of 86.2\% for dynamical stability of materials. A thermodynamic analysis of polymorphic stability in 126 systems demonstrates good agreement with DFT results at 300 K and 1000 K. In addition, the diverse and extensive high-quality DFT dataset curated in this study serves as a valuable resource for researchers to train and improve other machine learning interatomic potential models.


\end{abstract}
\maketitle

\section{\label{sec:introduction} Introduction}

Phonons, as quasiparticles representing the collective vibrational modes of atoms within a crystalline material, are ubiquitous and play a crucial role in determining various material properties, including thermal conductivity, mechanical behavior, electrical conductivity, and superconductivity \cite{kittel2018introduction, ziman2001electrons, wallace1998thermodynamics}. Additionally, they are vital in assessing dynamic and thermodynamic stability, as well as phase transitions of crystalline materials. This is particularly significant in materials discovery, where accurately predicting stability is essential for identifying new materials with desired properties \cite{griesemer2023accelerating, schmidt2017predicting, schleder2019exploring}.

One popular approach for first-principles phonon calculations is the finite-displacement method \cite{frank1995ab, Parlinski1997}. In this approach, the equilibrium positions of atoms are perturbed by small displacements, and the resulting changes in energies and forces are calculated to determine the force constants that govern the vibrational modes. However, it is worth noting that this method requires calculations of numerous supercells using density functional theory (DFT) \cite{kohn1965self, kohn1999nobel} to capture short to long-range interactions and achieve converged results. Consequently, phonon calculations are computationally intensive, especially for large unit cells or complex materials with low symmetry, and the resulting high-quality dataset is available for only a small set of materials \cite{togo_database, petretto2018high, okabe2023virtual}. Despite the exponential growth of computing power over the decades, traditional methods remain limited in their applicability to a vast array of materials in a high-throughput screening.

In recent years, machine learning approaches have emerged as a powerful tool for predicting phonon properties and accelerating materials discovery processes. These methods can be broadly categorized into two main strategies. The first strategy is directly predicting phonon properties using machine learning models trained on large datasets of phonon spectra. By leveraging advanced algorithms and techniques such as graph neural networks (GNN), these methods can predict phonon behaviors without constructing interatomic potentials \cite{gurunathan2023rapid, chen2021direct,okabe2023virtual,nguyen2022predicting}. For example, Gurunathan \etal \cite{gurunathan2023rapid} developed the atomistic line graph neural network (ALIGNN) which consists of crystal graph neural network and line graph including bond connectivity and bond angle information. By using the ALIGNN model, they achieved direct predictions of phonon density of states and other thermodynamic properties with good accuracy. Similarly, Chen \etal \cite{chen2021direct} demonstrated the direct prediction of phonon density of states using an Euclidean neural network (E(3)NN) \cite{geiger2022e3nn} which captures the symmetry of the crystal structures. It shows that the data-efficient E(3)NN model achieves reliable predictions with a small training dataset of 1,200 examples covering 64 types of elements. Additionally, Okabe \etal \cite{okabe2023virtual} used the virtual node graph neural network (VGNN) to directly predict $\Gamma$-phonon spectra and full dispersion without establishing an energy model through the training process. Their results show a significant speed-up in calculations while maintaining reliable accuracy. Furthermore, Nguyen \etal \cite{nguyen2022predicting} developed a deeper graph neural network with a global attention mechanism (deeperGATGNN), suggesting the potential of deeper graph neural networks in predicting phonon vibrational frequencies.

The second strategy is constructing machine learning interatomic potentials (MLIPs), also known as machine learning force fields, to predict phonon properties. In this approach, the machine learning training process aims to learn the functional relationship between crystal structures and potential energy surfaces (PES) without directly solving physical equations. At the early stage, kernel-based approaches and shallow neural networks were proposed, such as Behler-Parrinello neural network \cite{behler2007generalized} and Gaussian approximation potentials (GAP) \cite{bartok2010gaussian}. More recent methods have leveraged deep learning techniques, such as message-passing neural networks (MPNNs) \cite{MPNN}, and the models show remarkable accuracy \cite{SchNet, DimeNet, NequIP, M3GNet, MACE}.  Notably, the materials graph with three-body interactions neural network (M3GNet) \cite{M3GNet} demonstrated a potential for phonon predictions, showing promising results on averaged phonon frequencies across diverse materials in Materials Project \cite{jain2013commentary}. In our previous work \cite{MLUHIP}, we trained an improved version of directional message passing neural network, DimeNet++\cite{gasteiger2020fast}, by creating an intermediate force-displacement representation, which serves as a bridge between the existing phonon database, represented by interatomic force constants, and MLIP models. The trained model demonstrated accurate predictions of full harmonic phonon spectra and vibrational free energies. Rodriguez \etal \cite{rodriguez2023unlocking} developed an elemental spatial density neural network force field (Elemental-SDNNFF) for high-accuracy phonon property prediction of 77,091 structures. Although the model is limited to cubic materials, the researchers demonstrated an indirect machine learning approach that could effectively screen unexplored structures and identified 13,461 dynamically stable cubic structures with a lattice thermal conductivity below 1 W/m·K.

Despite significant progress of advanced neural networks and the remarkable performance of recent MLIP models, there remains a gap between predicted values from the models and experimental measurements in the real world. One of the primary factors contributing to this discrepancy is the quantity and quality of the phonon data used for training machine learning models. Currently, only a limited number of materials databases are available for phonons \cite{togo_database, petretto2018high, okabe2023virtual}. The most extensive phonon database known is the MDR phonon calculation database \cite{togo_database}, including full dispersion, projected density of states, and thermal properties of 10,034 compounds. Additionally, Petretto \etal \cite{petretto2018high} reported a database covering full phonon dispersion for 1,521 inorganic compounds, and Okabe \etal \cite{okabe2023virtual} built a machine learning-based $\Gamma$-phonon database comprising over 146,000 materials in Materials Project \cite{jain2013commentary}. While the MDR phonon calculation database covers an impressive number of compounds, it is still insufficient to accurately predict phonon properties for unknown materials, and other phonon databases cover only a subset of materials or are restricted to $\Gamma$-point. The challenge of constructing a large phonon database remains due to the intensive computational cost associated with phonon calculations.

In this study, we propose an alternative approach to accelerate harmonic phonon calculations by efficiently generating a training dataset. Leveraging a machine learning model, we significantly reduce the number of supercells requiring DFT self-consistent calculations. Instead of computing a large number of supercells with small displacements (typically 0.01~\r{A}) of a single atom, we generate a subset of supercell structures for each material, with all atoms randomly perturbed with displacements ranging from 0.01 to 0.05~\r{A}, where are typically used for extracting force constants using the Compressive Sensing Lattice Dynamics approach~\cite{csld}.  Through a preliminary analysis, we found that using only six structures for each material achieves a good balance between computational efficiency and prediction accuracy.  Subsequently, the resulting supercell structures and the interatomic forces obtained from DFT calculations, constitute the training dataset for our machine learning model. It is also worth noting that the results are expected to be systematically improvable with an increased number of training structures.

Our approach has two key features. First, by perturbing all atoms within supercells and increasing the amount of displacements, we gather numerous non-zero interatomic forces with relatively large magnitudes and rich information. Second, we use a data-driven approach to compensate for the reduced number of supercells. The underlying reasoning is that some materials may share common structural features, such as identical types of elements or similar bonding environments. For example, even when two materials have entirely different elemental compositions, certain structural features, such as the radii or electronegativities of ions, could show similarities. We hypothesize that if we train a machine learning model on a diverse range of materials, the model can identify such underlying similarities across different structures by itself. Leveraging the capability of existing architectures of machine learning interatomic potentials, we can significantly reduce the requisite number of supercells for phonon calculations. This approach enables us to construct training datasets using only a subset of supercells per material while maintaining high accuracy in predicting harmonic phonon properties.

This paper is organized as follows. In Section \ref{sec:method}, we introduce the machine learning model, MACE \cite{MACE}, used in this study, and provide details of DFT and harmonic phonon calculations. In Section \ref{sec:results}, we provide details of the training dataset we construct, and demonstrate the performance of the model in predicting harmonic phonon properties (vibrational frequencies, full phonon dispersions, and Helmholtz vibrational free energies), including dynamic stability, and thermodynamic stability of materials. Finally, in Section \ref{sec:conclusion}, we summarize the key findings of our study, including the trained MACE machine learning model performance, and discuss the limitations of our approach and the potential implications for further research in this field.

\section{\label{sec:method} Method}


\subsection{MACE}

MACE \cite{MACE} is a state-of-the-art framework for MLIPs with not only high accuracy but also high computational efficiency. MACE was built based on MPNNs \cite{MPNN}, a type of GNNs. GNNs represent an atomic structure as a graph using nodes and edges: a node corresponds to an atom, and an edge connects two nodes if the distance between them is less than a cut-off radius, $r_{cut}$. In the message passing phase of the MPNNs, messages are generated by pooling over the states of neighboring nodes and sent to a target node as shown in Fig.\ref{fig:training} (a), and the atomic features of the target node are updated. After several message passing iterations, the updated atomic states are mapped onto atomic site energies in the readout phase. The total potential energy of a crystal structure can be obtained by summing the site energies, and the forces and the stresses of the structure are typically derived using auto-differentiation techniques on the total potential energy \cite{M3GNet, MACE_evaluation}.
Early MPNN models such as SchNet \cite{SchNet} and DimeNet \cite{DimeNet} used invariant messages under rotations of the atomic structure. Subsequently, equivariant MPNN models using geometric tensors were developed to achieve data efficiency and high accuracy \cite{Tensor_field, Cormorant, NequIP}. However, as most MPNN models use only two-body messages, a large number of message passing iterations are required for high accuracy, leading to the relatively high computational cost \cite{MACE}. 

The key idea of the MACE model is constructing high-body-order equivariant messages in each layer of the MPNNs by applying Atomic Cluster Expansion (ACE) \cite{ACE}. The increased body-order of messages reduces the number of message passing iterations (number of layers, $S$) required to converge in accuracy, thereby increasing calculation speed. It was demonstrated that only two layers are required when using the MACE model with 4-body messages, while other MPNN models need 4 to 6 layers with two-body messages for high accuracy \cite{MACE}. Fig.\ref{fig:training} (a) shows two layers of the MACE model with 4-body messages (corresponding to three correlation orders, $\nu=3$) in each layer. At the first message passing iteration, the state of a target node is updated using messages formed by its three nearest-neighbor nodes. In the second iteration, next-nearest-neighbor nodes are involved in the message passing process, expanding the region used to determine the site energy to $2r_{\text{cut}}$. Hence, the MACE model with two layers covers a total 13-body features. The many-body order messages are efficiently built through tensor products carried out on the nodes \cite{MACE, MACE_design}. This is a unique feature of the MACE architecture, making the model highly parallelizable and fast. For details of the MACE architecture, see the references \cite{MACE, MACE_design, MACE_evaluation}.

In this research, we used a MACE model with two interaction layers ($S=2$), three correlation orders ($\nu=3$), and a cut-off radius ($r_{\text{cut}}$) of 6~\r{A} for each layer. The number of embedding channels ($k$) and the maximum of the symmetry order, $L_\mathrm{max}$, of the messages were set to 64 and 1, respectively, representing 64 equivariant messages. These two hyperparameters, $k$ and $L_\mathrm{max}$, mainly determine the model size \cite{MACE_evaluation}. While higher accuracy can be achieved by increasing the number of layers and including higher-order features such as matrices and tensors with additional embedding channels, our choices were made with careful consideration of computational costs. Reference energies were estimated using least square regression by using the average option ($E
_\text{0s}=\text{average}$). Both batch size and valid batch size were set to 4, and the random seed was fixed at 123. 95\% of our training dataset was used for training, while the remaining 5\% was allocated for validation. We set a force weight of 1000 in the loss function, without energy weight. For the details of the loss function, see Ref.\cite{MACE_evaluation}. By setting the energy weight to zero, only forces were used to train the model, and it enabled the model to predict forces very accurately. It was demonstrated by D\'{a}vid \textit{et al.}~\cite{MACE_evaluation} using a high loss weight on the forces compared to other properties results in the most accurate models, especially when the training dataset covers diverse types of materials. After 100 epochs, we reduced the weight factors on the forces from 1000 to 100, and the model was trained until 200 epochs. We also conducted several tests by changing the values of some hyperparameters, as detailed in Appendix \ref{sec:appendix}. For other hyperparameters not mentioned in this paper, we used default settings. The efficiency of the model was demonstrated by the low computational cost for the training, which is only about 72.3 GPU hours using the
NVIDIA Tesla V100 32GB GPU of Bridges-2 \cite{brown2021bridges}. 

\subsection{DFT calculations}
DFT \cite{dft} calculations were conducted using Vienna Ab initio Simulation Package (VASP) \cite{Vasp1, Vasp2, Vasp3, Vasp4}. The projector augmented wave (PAW) \cite{paw} method was used with the Perdew-Burke-Ernzerhof (PBE) \cite{pbe,Perdew2008} of generalized gradient approximation (GGA) \cite{gga} exchange-correlation functional. A plain wave basis set was used with a cutoff energy of 520 eV. We conducted structure relaxation calculations for 2,738 crystal structures using DFT. The criteria of energy and force convergence were set to $10^{-8}$ eV and $10^{-3}$ eV/\r{A}, respectively, and the number of k-points was determined by setting the parameter KSPACING of 0.15. After that, we generated supercell structures containing approximately 100-200 atoms based on the relaxed structures and randomly perturbed all atoms with an amount of 0.01-0.05\r{A}. These perturbations induce non-zero forces between atoms as the displaced atoms deviate from their equilibrium positions. For each distinct structure, we generated approximately six perturbed structures, resulting in a total 15,670 structures, and performed DFT self-consistent calculations to calculate interatomic forces. All of the perturbed structures and their corresponding force sets calculated from DFT were used as a training dataset of the MACE model. Due to the high computational cost and limited resources available, we generated a harmonic phonon dataset by downselecting 384 structures among the 2,738 structures.  To create the phonon dataset, we employed the finite-displacement method using the Phonopy package\cite{Phonopy_first, Phonopy_implementation}, and performed DFT self-consistent calculations with an energy convergence threshold of $10^{-8}$ eV. To ensure the high-quality of phonon dataset, we used a dense k-points mesh with a KSPACING parameter set to 0.15 in the DFT self-consistent calculations. The phonon dataset serves as a benchmark to evaluate the performance of the trained MACE machine learning model in predicting harmonic phonon properties.

\subsection{Harmonic phonons}

The potential energy ($E_p$) of a crystal structure can be expanded in a Taylor series:
\begin{equation} \label{eq1}
\begin{split}
E_p & = \Phi_0 + \sum_{l\kappa\alpha} \Phi_{\alpha}^{l\kappa} \textbf{u}_{\alpha}^{l\kappa} \\
& + \frac{1}{2!} \sum_{l\kappa\alpha, l'\kappa'\alpha'} \Phi_{\alpha,\alpha'}^{l\kappa, l'\kappa'} \textbf{u}_{\alpha}^{l\kappa} \textbf{u}_{\alpha'}^{l'\kappa'}\\
& + \frac{1}{3!} \sum_{l\kappa\alpha, l'\kappa'\alpha', l''\kappa''\alpha''} \Phi_{\alpha,\alpha',\alpha''}^{l\kappa, l'\kappa', l''\kappa''} \textbf{u}_{\alpha}^{l\kappa} \textbf{u}_{\alpha'}^{l'\kappa'} \textbf{u}_{\alpha''}^{l''\kappa''} + ...
\end{split}
\end{equation}
where $l, l', ...$ and $\kappa, \kappa', ...$ are the indices of unit cells and the atoms of the corresponding unit cell, respectively, and $\alpha, \alpha', ...$ denote the Cartesian directions (x, y, z). $\textbf{u}_{\alpha}^{l\kappa}$ represents the displacement of the atom $\kappa$ of the unit cell $l$ along the $\alpha$ direction from its equilibrium position. $\Phi_0, \Phi_{\alpha}^{l\kappa}, \Phi_{\alpha,\alpha'}^{l\kappa, l'\kappa'}, \Phi_{\alpha,\alpha',\alpha''}^{l\kappa, l'\kappa', l''\kappa''}, ... $ are the zeroth-, first-, second-, third-, ... order of interatomic force constants (IFCs), respectively. 

The first term, $\Phi_0$, is a constant value of energy shift that can be chosen as a reference. It is assumed that $\Phi_0 = 0$. Additionally, we set $\Phi_{\alpha}^{l\kappa} = 0$ because the Taylor series is conducted around the equilibrium position. Under the harmonic approximation, we neglect higher orders such as third, fourth, and so on, in the Eq.\ref{eq1}, considering only small displacements from the equilibrium position. Consequently, the potential energy under the harmonic phonon approximation is simplified as:
\begin{equation} \label{eq2}
E_p = \frac{1}{2} \sum_{l\kappa\alpha, l'\kappa'\alpha'} \Phi_{\alpha,\alpha'}^{l\kappa, l'\kappa'} \textbf{u}_{\alpha}^{l\kappa} \textbf{u}_{\alpha'}^{l'\kappa'}
\end{equation}

With an atomic force $F_{\alpha}^{l\kappa} = - \partial E_p / \partial \textbf{u}_{\alpha}^{l\kappa}$, the second-order IFC is written as:
\begin{equation} \label{eq3}
\Phi_{\alpha,\alpha'}^{l\kappa, l'\kappa'} = \frac{\partial^2 E_p}{\partial \textbf{u}_{\alpha}^{l\kappa} \partial \textbf{u}_{\alpha'}^{l'\kappa'}} = - \frac{\partial F_{\alpha'}^{l'\kappa'}}{\partial \textbf{u}_{\alpha}^{l\kappa}} 
\end{equation}

For harmonic phonon calculations, dynamical matrix is constructed to solve an eigenvalue problem as follows \cite{Dove1993}:
\begin{equation} \label{eq4}
\sum_{\kappa' \alpha'} D_{\alpha \alpha'}^{\kappa \kappa'}(\textbf{q}) \textbf{e}_{\textbf{q} \nu}= \omega_{\textbf{q} \nu}^2 \textbf{e}_{\textbf{q} \nu}
\end{equation}
where $\textbf{q}$ is the wave vector, and $\nu$ is the phonon band index. $\textbf{e}_{\textbf{q} \nu}$ and $\omega_{\textbf{q} \nu}$ are the polarization vector, and the phonon frequency of each phonon mode $\textbf{q} \nu$, respectively. And the dynamical matrix $D_{\alpha \alpha'}^{\kappa \kappa'}(\textbf{q})$ is defined as:
\begin{equation} \label{eq5}
D_{\alpha \alpha'}^{\kappa \kappa'}(\textbf{q}) = \frac{1}{\sqrt{m_\kappa m_\kappa'}} \sum_{l'} \Phi_{\alpha,\alpha'}^{0\kappa, l'\kappa'} e^{i \textbf{q} (\textbf{r}_{l' \kappa'} - \textbf{r}_{0 \kappa'})}
\end{equation}
where $m_\kappa$ is the mass of the atom $\kappa$, and $\textbf{r}_{l' \kappa'}$ is the position of the atom $\kappa'$ in the unit cell $l'$. In this equation, the unit cell $l$ is indexed based on a reference ($l=0$), as indicated by $\Phi_{\alpha,\alpha'}^{0\kappa, l'\kappa'}$ and $\textbf{r}_{0 \kappa'}$.

If phonon frequencies are determined across the Brillouin zone, thermal properties such as Helmholtz vibrational free energy ($A_{\text{vib}}$), heat capacity at constant volume, and vibrational entropy can be calculated using the canonical ensemble in statistical mechanics under the harmonic approximation. Helmholtz vibrational free energy is calculated using the following equations \cite{Dove1993, Phonopy_first}:

\begin{equation} \label{eq6}
A_{\text{vib}} = \frac{1}{2} \sum_{\textbf{q} \nu} \hbar \omega_{\textbf{q} \nu} + k_BT \sum_{\textbf{q} \nu} \ln \left[1 - \exp \left(\frac{-\hbar \omega_{\textbf{q} \nu}}{k_BT}\right) \right]
\end{equation}
where $\hbar$, $k_B$, and $T$ are the reduced Plank constant, the Boltzmann constant, and the absolute temperature, respectively.

Following the above equations, it is clear that the second-order IFCs are required for the harmonic phonon calculations, and these IFCs can be calculated from the forces. Here, we generated two types of force sets on the phonon dataset: one calculated from DFT, and the other predicted from the trained MACE model. Using these force sets, corresponding IFCs were calculated. Subsequently, phonon dynamic and thermal properties such as phonon dispersion, Helmholtz vibrational free energies, constant volume heat capacities, and vibrational entropies were computed to evaluate the performance of the trained MACE model and compared to DFT.

All phonon calculations in this study were performed using the Phonopy package~\cite{Phonopy_first, Phonopy_implementation}. For both force sets obtained from DFT and MACE, we used the same supercell matrix for each structure. The choice of the supercell matrix was based on the size of the structures, typically containing 100-200 atoms in a supercell. In cases where pure elements consisted of only one atom in their primitive cell, a 4 $\times$ 4 $\times$ 4 supercell size was used. Helmholtz vibrational free energy, heat capacity at constant volume, and vibrational entropy were calculated at 300 K and 1000 K. These temperatures were chosen to assess the thermal dynamical stability of the materials, considering that some materials can be synthesized at high temperatures. Synthesis temperatures can vary depending on the materials being synthesized, synthesis methods, and desired properties. For the thermal property calculations, the $\textbf{q}$-point mesh was sampled by the parameter $l_q$ as a length, and the mesh numbers $N_1$, $N_2$, and $N_3$ along the reciprocal lattice vectors $b^*_1$, $b^*_2$, and $b^*_3$ are defined as $N_i = \text{max}[1, \text{nint}(l_q|b^*_i|)]$. Here, we used $l_q=75$ for the $\textbf{q}$-point mesh sampling, as implemented within Phonopy~\cite{Phonopy_first, Phonopy_implementation}. The cut-off frequency was set to 0.1 THz, excluding phonon modes with frequencies below this threshold in the thermal property calculations. Such a choice is based on an observation that when phonon frequencies are very small or close to zero, the phonon population based on the Bose-Einstein statistics increases exponentially, which can lead to both unphysical results and reduced computational efficiency.

\begin{figure*}[htp]
	\includegraphics[width = 0.9\linewidth]{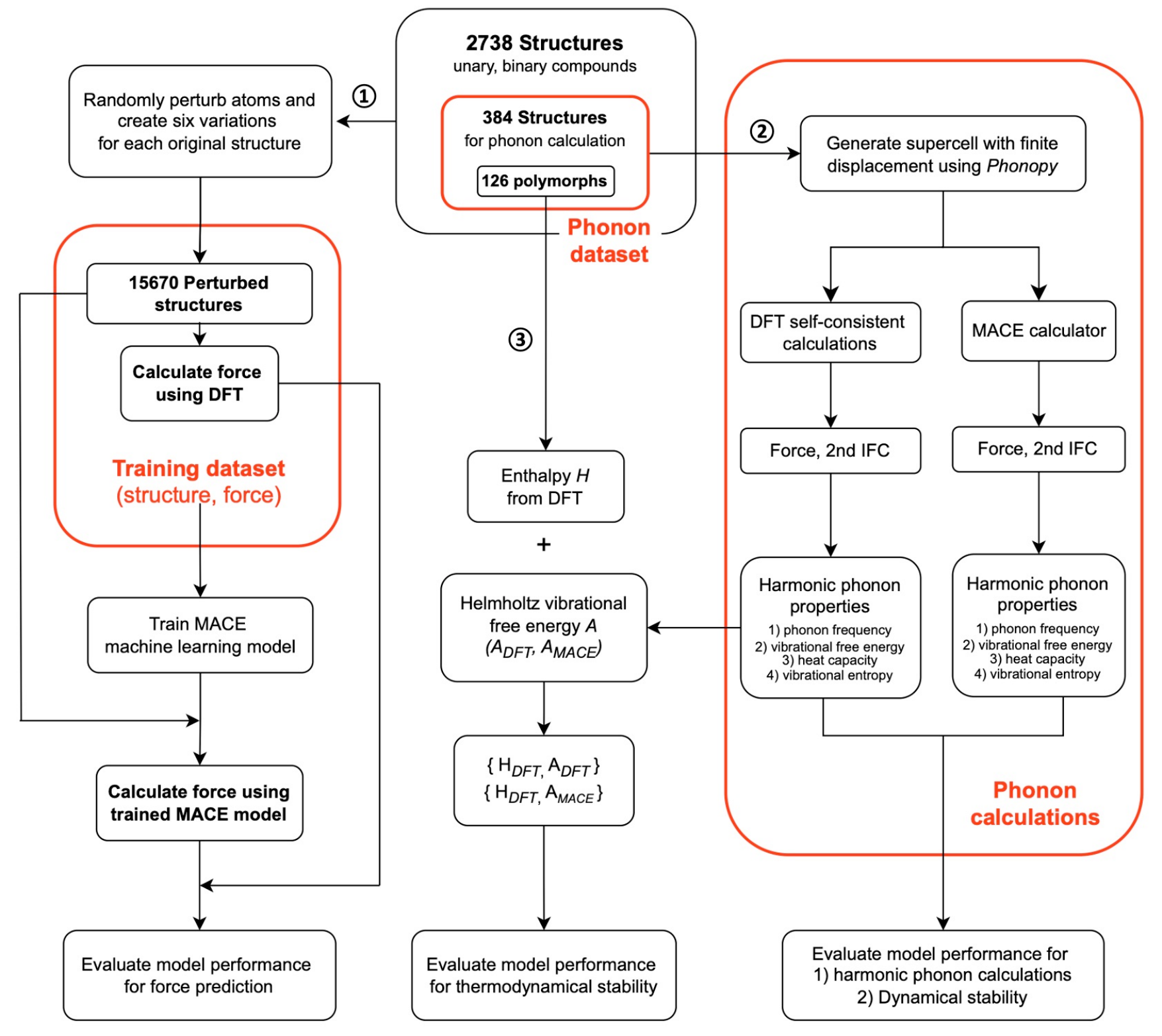}
	\caption{ 
	Workflow chart showing the computational processes employed in our study. In the initial step, we construct a training data set for the MACE machine learning model and evaluate its performance in predicting forces. Subsequently, we generate a phonon dataset consisting of 384 structures and calculate the harmonic phonon properties using both DFT and the trained MACE model. Comparative analysis between DFT and the MACE model is conducted to assess the performance of harmonic phonon predictions and the dynamical stability of materials. In the final step, we evaluate the performance of the trained MACE model in predicting thermodynamic stability and polymorphic transitions across 126 polymorphs, using enthalpy and Helmholtz vibrational free energy.
	}
	\label{fig:workflow}
\end{figure*}

Fig.~\ref{fig:workflow} provides a comprehensive overview of the workflow employed in this study. Initially, we construct a training dataset by perturbing all atoms within supercells and computing forces through DFT calculations. Next, we generate a phonon dataset comprising 384 structures and compute the harmonic phonon properties to evaluate the performance of the trained MACE machine learning model. Finally, we evaluate the performance of the model in predicting thermodynamic stability and phase transitions of polymorphs based on the enthalpy and Helmholtz vibrational free energies. The details of the training dataset and the performance of the machine learning model will be discussed in the following sections.

\section{\label{sec:results} Results and discussion}


\subsection{Analysis of training dataset}

\begin{figure*}[htp]
	\includegraphics[width = 0.85\linewidth]{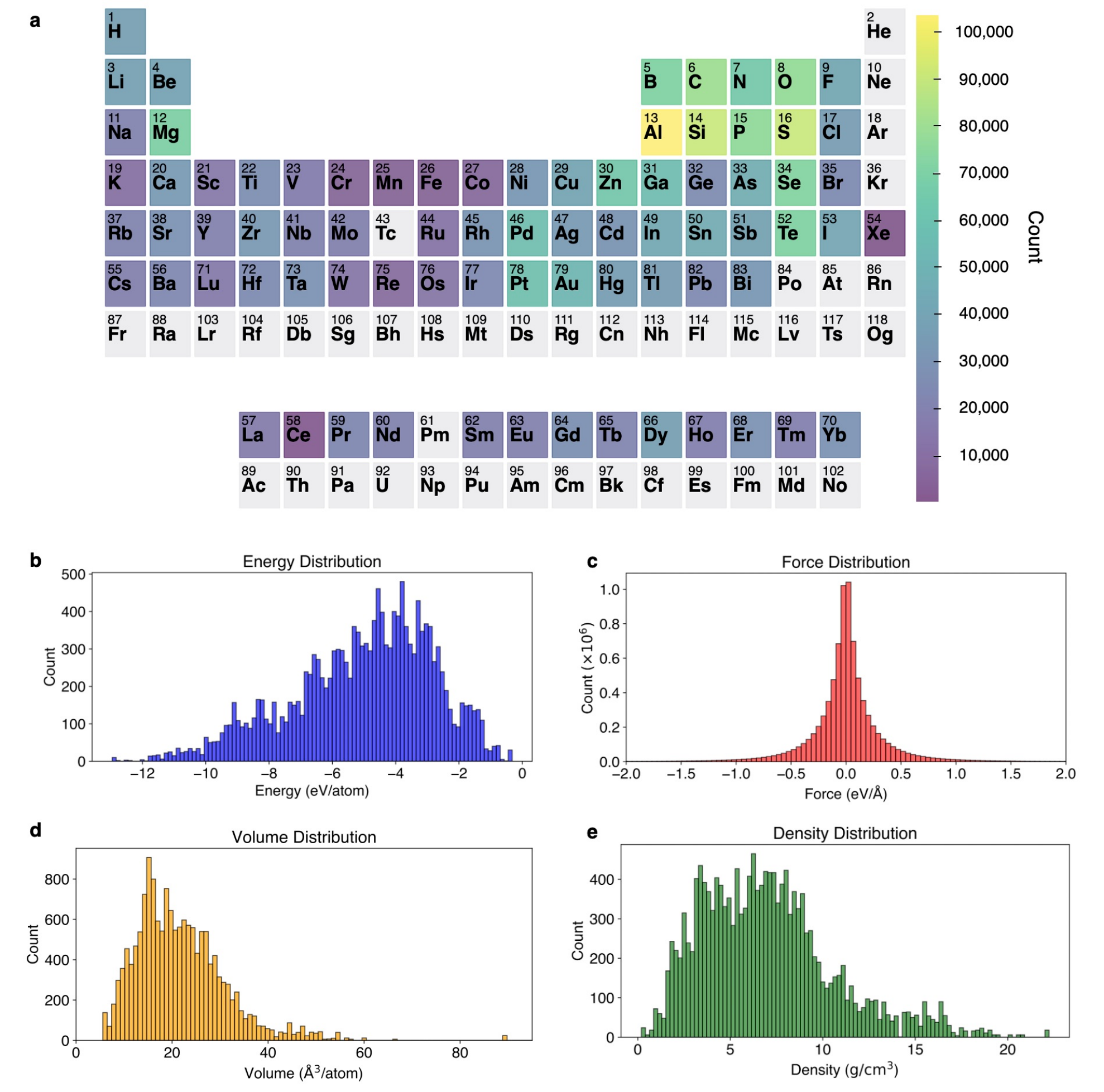}
	\caption{ 
	(a) The heat map of element counts for all atoms in the dataset covering a total 77 elements with atomic numbers ranging from 1 to 83. The elements with a grey color in the heat map are not included in the dataset. (b)-(e) The distribution of the dataset: (b) energy, (c) force, (d) volume, and (e) density distribution, respectively.
	}
	\label{fig:dataset}
\end{figure*}

The training dataset used in this study for the MACE model contains 15,670 crystal structures and 8.1 million force components of the structures. The dataset comprises pure elements and binary compounds. It covers 77 kinds of elements across the periodic table in the range of atomic number from 1 to 83, and few elements such as noble gases and radioactive elements are excluded. A heat map of the element counts for all atoms in the dataset is shown in Fig.\ref{fig:dataset} (a). Aluminum (Al) is the most frequently occurring element, with over 100,000 counts, while nearly all other elements are represented tens of thousands of times. It indicates that the training dataset contains a diverse range of pure elements and binary compound materials. Only two elements, Ce and Xe, have lower counts than 1,000. Fig.\ref{fig:dataset} (b)-(e) show the distribution of the dataset in terms of the energy (eV/atom), force (eV/\r{A}), volume ($\mathrm{\AA}^3$/atom), and density ($\mathrm{g/cm^3}$), respectively. Energies, forces, volumes, and densities are in range of [-12.96, -0.32] eV/atom, [-40.73, 40.73] eV/\r{A}, [5.57, 142.83] $\mathrm{\AA}^3$/atom, and [0.19, 22.23] $\mathrm{g/cm^3}$, respectively. As shown in Fig.\ref{fig:dataset} (c), the force distribution is symmetric, and the mean and the standard deviation of force are 0.0 eV/\r{A} and 0.534 eV/\r{A}, respectively.

\subsection{Assessment of machine learning models}

\begin{figure*}[htp]
	\includegraphics[width = 0.95\linewidth]{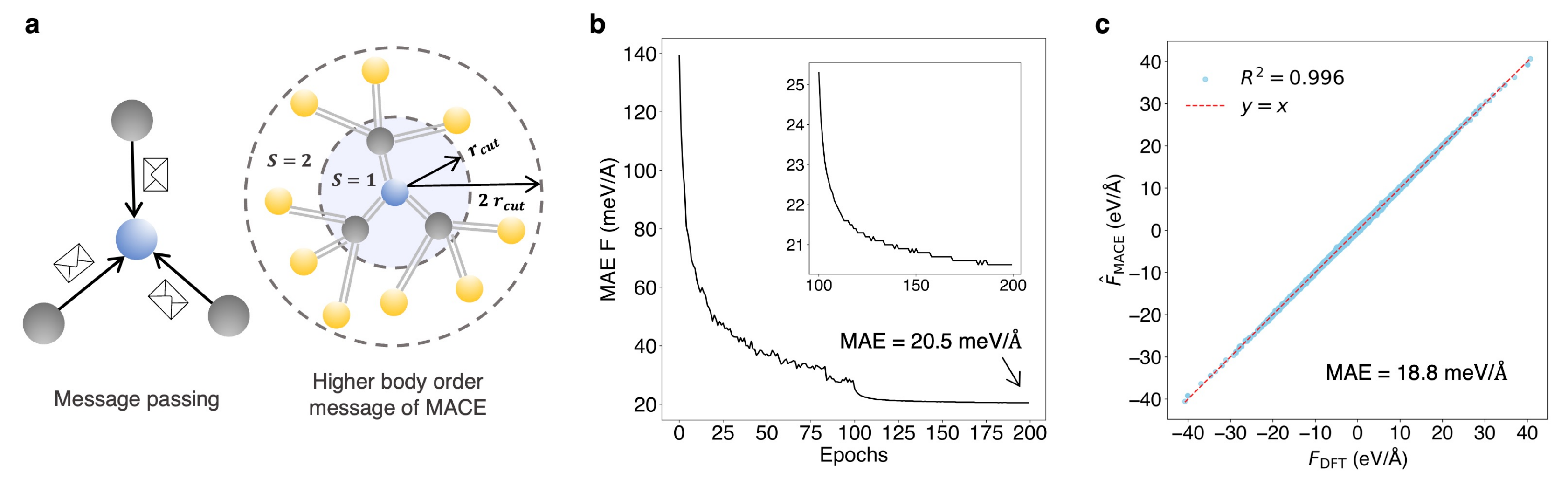}
	\caption{ 
	(a) The schematic diagrams of message passing (left) and the higher body order message construction of MACE with two layers (right). (b) The force validation errors of the MACE model during training. The inset plot shows the decrease in validation error from 100 to 200 epochs. (c) The scatter plot represents predicted (MACE) versus calculated (DFT) values for all force components in the training dataset. The dashed red line indicates the 1:1 correlation.
	}
	\label{fig:training}
\end{figure*}

To evaluate the performance of the trained MACE model, we use the mean absolute error (MAE) and the coefficient of determination ($R^2$). They are defined as follows:
\begin{equation} \label{eq8}
\text{MAE} = \frac{1}{n} \sum_{i=1}^{n}|y_{i} - \hat{y}_{i}|
\end{equation}
\begin{equation} \label{eq9}
R^2 = 1 - \frac{\sum_{i=1}^{n}(y_{i}-\hat{y}_{i})^2}{\sum_{i=1}^{n}(y_{i}-\Bar{y})^2}
\end{equation}
where n is the number of data points, $y_{i}$ and $\hat{y}_{i}$ are the actual and predicted value of the $i$th data, respectively. $\Bar{y}$ indicates the mean of actual values in a dataset. The $R^2$ value measures the proportion of variation in the predicted values that can be predicted from the actual values, ranging from 0 to 1. A higher $R^2$ value indicates better prediction performance of a model.

We used 5\% of the training dataset for validation, and the learning curve during the training is shown in Fig.\ref{fig:training} (b). The MAE of forces for the validation dataset consistently decreases, reaching 20.5 meV/\r{A} after 200 epochs. We compared the predicted values for all force components in the training dataset with the corresponding values calculated using DFT, as shown in Fig.\ref{fig:training} (c). The MAE of forces for all training dataset points is 18.8 meV/\r{A} with the 8.60 \% of relative force MAE, and the $R^2$ is 0.996. These results show that our trained MACE model achieved a significantly lower MAE for force prediction, compared to the previous models such as the materials graph with three-body interactions neural network (M3GNet) \cite{M3GNet}, the graph-based pre-trained transformer force field (GPTFF) \cite{xie2024gptff}, and the machine learning universal harmonic interatomic potential (MLUHIP) \cite{MLUHIP}; the MAE values on the phonon dataset were reported as 72 meV/\r{A} (M3GNet), 71 meV/\r{A} (GPTFF), and 78 meV/\r{A} (MLUHIP), respectively \cite{M3GNet, xie2024gptff, MLUHIP}. We acknowledge that the MAEs cannot be compared directly because the reported MAE values of the previous models were calculated on the test dataset, while our 18.8 meV/\r{A} of MAE was calculated from the training dataset. For the validation dataset, the MAE was 20.5 meV/\r{A} which is larger than the MAE of the whole training dataset.

We examined if our trained MACE model based on only forces (MACE-F) could be enhanced by additional training with energies. Since the MACE-F model has not been trained on energies, the energy MAE of the model is 467.2 meV/atom on the validation set. We conducted an additional 100 epochs of training on the MACE-F model including both force and energy contributions in the loss function with weights of 100 and 50, respectively. It was a total 300 epochs of training (200 epochs in the first phase with forces + 100 epochs in the second phase with forces and energies). The same hyperparameters were used, except for the force and energy loss weights. After the second-phase training, the model trained with forces and energies (MACE-FE) shows an energy MAE of 10.5 meV/atom and a force MAE of 23.8 meV/\r{A}. The energy MAE is significantly decreased from 467.2 meV/atom to 10.5 meV/atom, while the force MAE, starting at 57.1 meV/\r{A} in the initial epoch of the second phase (i.e., after 201 epochs), consistently decreases, reaching 23.8 meV/\r{A} at the end. Even though the force MAE continuously decreases during the second-phase training, it is still higher than the MAE of 20.5 meV/\r{A} of the MACE-F model. It is worth noting that the higher force MAE leads to a lower accuracy of predictions for the harmonic phonon properties. We also conducted different training on the MACE-F model by changing the loss weight ratio of the second phase. With the force weight of 100 and the energy weight of 10, the energy MAE is 35.3 meV/atom on the validation set, which is higher than the MACE-FE model, while the force MAE is barely changed 23.2 meV/\r{A}. When the energy weight is higher than the force weight in training, the energy prediction is very accurate (the MAE is less than 8.5 meV/atom), but the force error is significantly increased compared to those of the MACE-F and MACE-FE.

Furthermore, in the initial stage, we conducted several training tests on the effects of the force and energy loss weight ratio, starting from the vanilla MACE model. We found that the force error becomes the lowest when energies are \textit{not} used in the initial training. For details of the training results, see Appendix \ref{sec:appendix}. Therefore, we conclude that training with energy is not necessary for very accurate force predictions in the case of our study. As the force and energy weights in the loss function are crucial to the results, those weights should be chosen carefully \cite{MACE_evaluation}. Based on these findings, we confirmed that our MACE-F model which is trained only with forces is the optimal choice for force and phonon property predictions. In the following sections, we will discuss the performance of the MACE-F model on predicting dynamical and thermodynamic stability based on phonon calculations.

\subsection{Phonon frequency and dynamical stability}

\begin{figure*}[htp]
	\includegraphics[width = 0.85\linewidth]{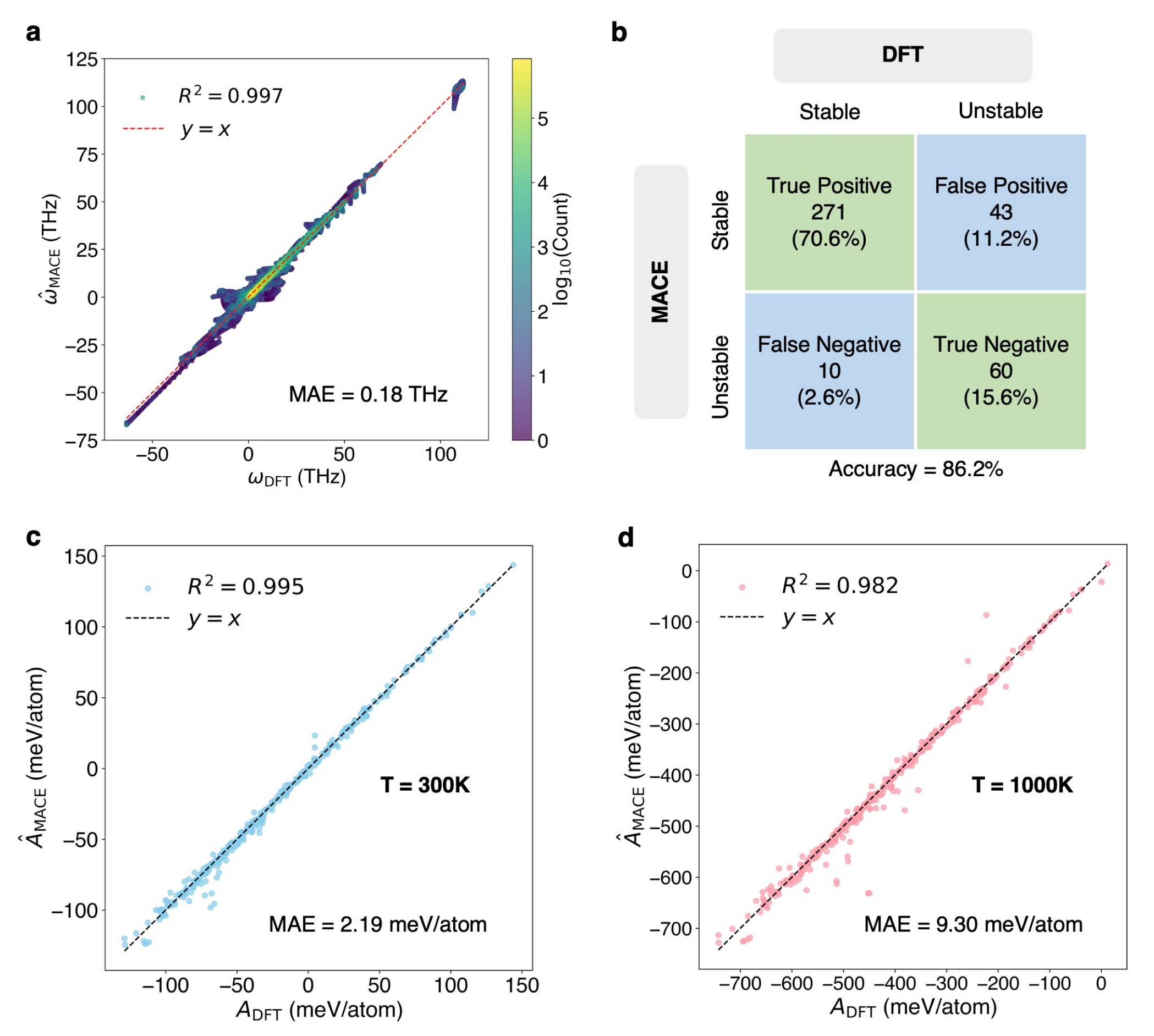}
	\caption{ 
	Evaluation of the trained MACE model performance on a phonon dataset comprising 384 materials. (a) The scatter points represent predicted (MACE) versus calculated (DFT) phonon frequencies obtained along the high-symmetry path of each crystal structure. The distribution of scatter points is visualized using a two-dimensional histogram with a log scale. Bright-color regions indicate a high concentration of data points. (b) Confusion matrix for dynamical stability predictions. The accuracy of dynamical stability prediction is 86.2\%, where the tolerance parameter for counting imaginary phonon modes was set to 0.5 THz. (c) and (d) show the comparison of predicted Helmholtz free energies using our trained MACE model with DFT-calculated values at (c) 300 K and (d) 1000 K, respectively. 
	}
	\label{fig:evaluation}
\end{figure*}

In our analysis of the MACE-F model performance in predicting dynamical stability, we used a phonon dataset consisting of 384 materials. The size of the phonon dataset may not be extensive due to the high computational costs associated with phonon calculations. Generating a large, high-quality phonon dataset presents challenges as it requires numerous calculations with dense k-points. Nevertheless, the phonon dataset we constructed is not restricted to specific studies or materials, covering diverse elements. Thus, we believe this phonon dataset enables the assessment of our model performance as a universal machine learning interatomic potential.

We compared the phonon frequencies predicted by the MACE model and those obtained from DFT. As mentioned in Section \ref{sec:method}, the second-order IFCs are necessary for harmonic phonon calculations, which are derived from force sets. Based on the predicted forces using the MACE model, we computed full phonon dispersions for structures in the phonon dataset along the high-symmetry path in the reciprocal space. Similarly, DFT-calculated phonon dispersions were obtained. A comparison of the phonon dispersions between MACE and DFT is provided in Supplementary Material. Fig.\ref{fig:evaluation} (a) presents a scatter plot representing a total of 7.8 million phonon frequency (THz) data points, plotted as (DFT frequency, MACE frequency) pairs. The distribution of these data points is visualized using a two-dimensional histogram, with a color bar indicating the concentration of the data points on a logarithmic scale. While there are variations of the frequencies between MACE and DFT, particularly near zero, it should be noted that the number of data points is around 7.8 million, and 95\% of them are within the range of [0.50, 29.6] THz. In Fig.\ref{fig:evaluation} (a), the data points marked as bright colors follow the guideline ($y=x$), indicating most of the predicted values are in good agreement with the calculated values. Additionally, the MAE of the phonon frequency is only about 0.18 THz, with the $R_2$ value of 0.997. The error is significantly smaller compared to other machine learning models \cite{nguyen2022predicting, M3GNet, MLUHIP}. Even though we trained the MACE model using fewer than six supercells for each structure and the phonon dataset covers a diverse range of materials, the trained MACE model demonstrates highly accurate performance in predicting phonon frequencies. It demonstrates that we can effectively reduce the number of supercells requiring DFT self-consistent calculations for generating a training dataset of phonon machine learning models while still obtaining accurate phonon frequency predictions through the training process.

The presence of imaginary phonon modes generally represents the dynamical instability of a crystal structure, and the determination of the dynamic stability of structures is essential in the material discovery field. The dynamical stability analysis for the phonon dataset was conducted using phonon frequencies calculated from both MACE and DFT. Fig.\ref{fig:evaluation} (b) illustrates the confusion matrix of the dynamical stability predictions for the 384 materials in the phonon dataset. The tolerance for determining dynamic instability was set to 0.5 THz considering the possibility of numerical errors, meaning that structures were classified as unstable if any phonon frequency of the structure was lower than -0.5 THz. The trained MACE model achieves 86.2\% accuracy on the dynamical stability classification with true negative cases of 15.6\%. Although there are 11.2\% false positive cases, false negative cases are only 2.6\%. It suggests that our trained MACE model can be potentially used as an initial filter for high-throughput screening of dynamically unstable materials.

\subsection{Thermodynamic stability of polymorphs}

\begin{figure*}[htp]
	\includegraphics[width = 0.85\linewidth]{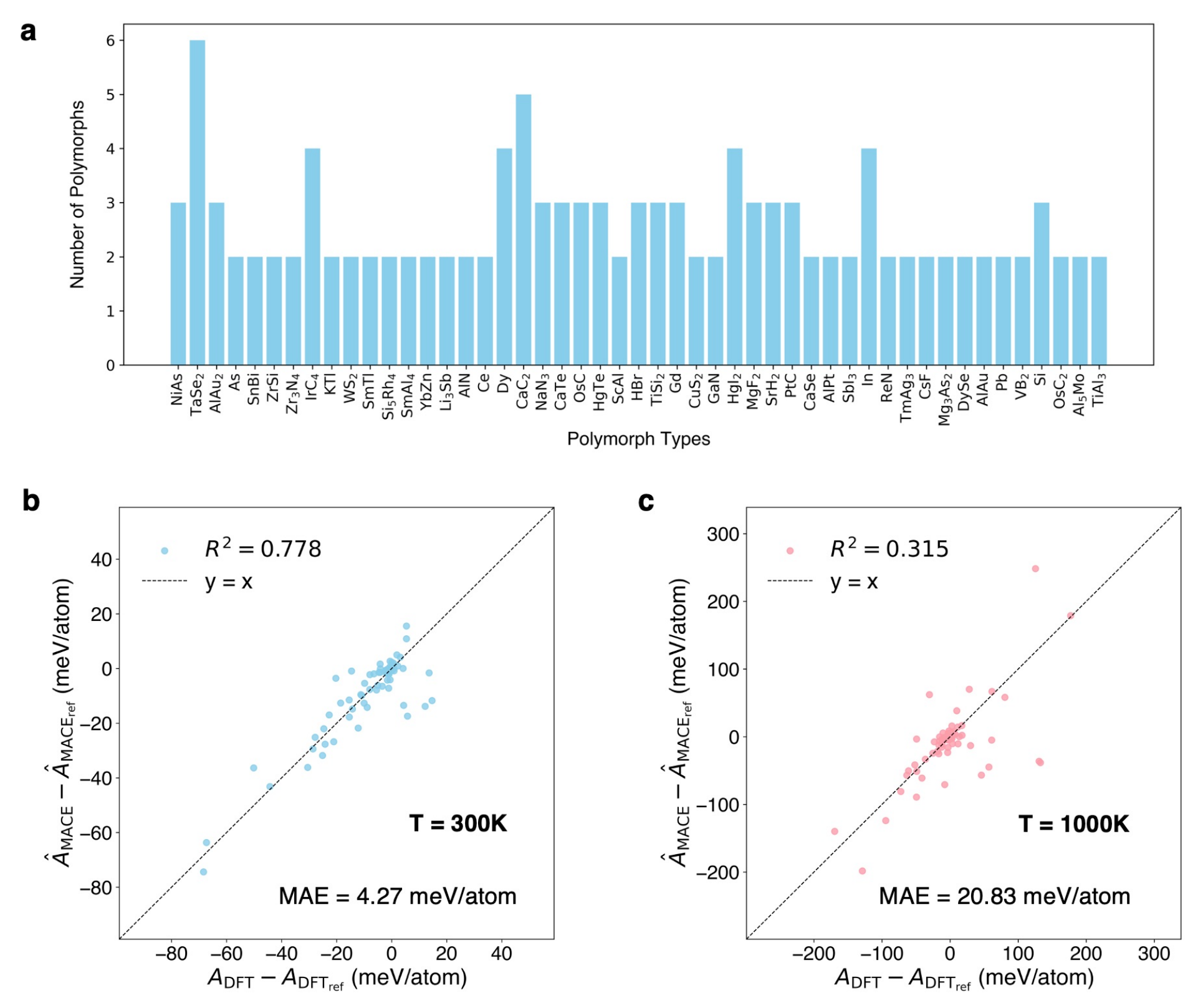}
	\caption{ 
 	(a) The histogram shows 49 distinct polymorph types in the phonon dataset, with the x-axis representing each type and the y-axis indicating the number of polymorphs. The phonon dataset contains a total of 126 polymorphs distributed across these 49 types. (b) and (c) show the difference in Helmholtz free energies between each polymorph within a specific polymorph type and its corresponding reference structure. The scatter plots the predicted energy differences derived from the trained MACE model, compared with energy differences calculated using DFT at (c) 300 K and (d) 1000 K, respectively.
	}
	\label{fig:polymorph}
\end{figure*}

\begin{figure*}[htp]
	\includegraphics[width = 0.85\linewidth]{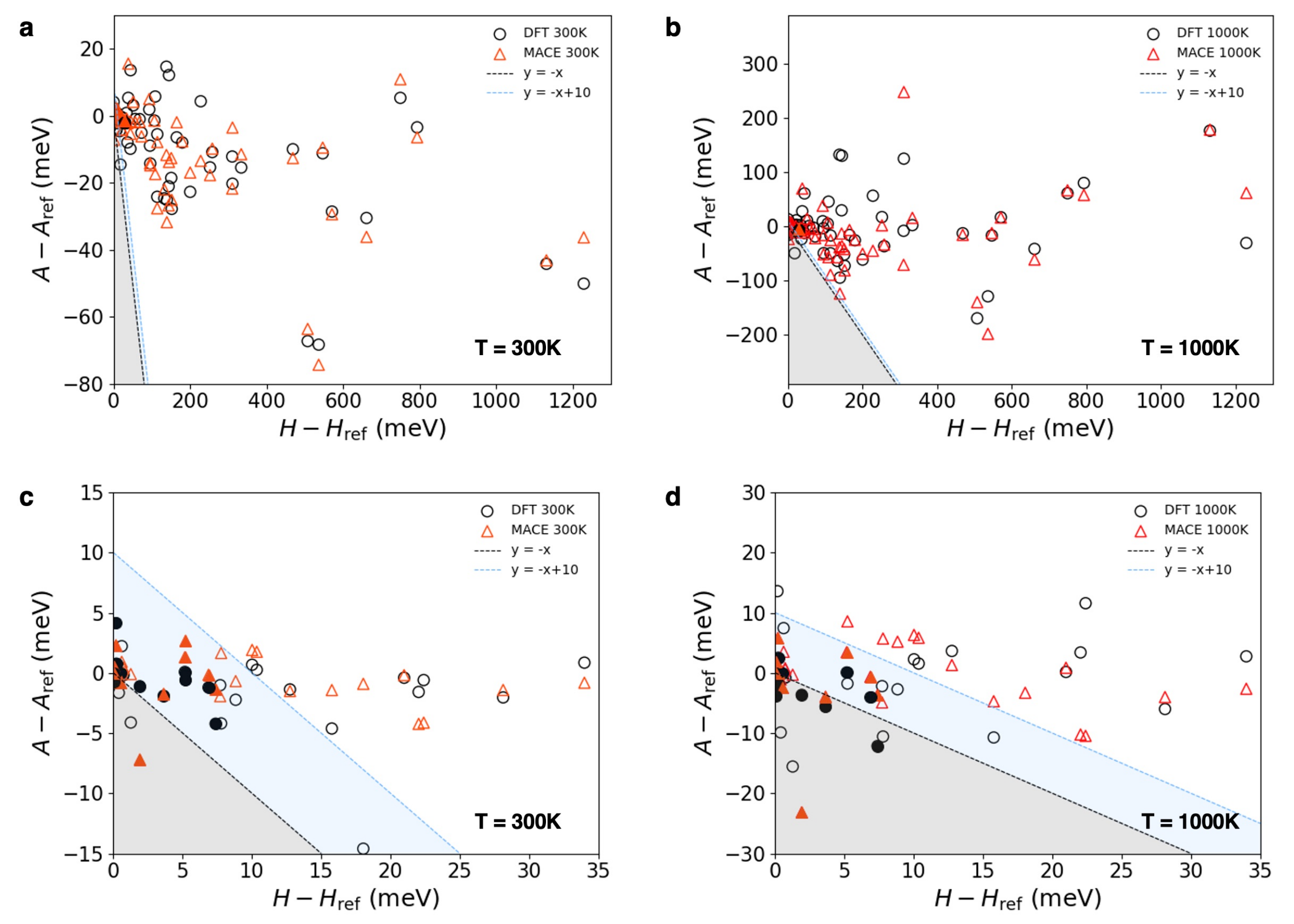}
	\caption{ 
    In the scatter plots (a)-(d), the x-axis denotes the enthalpy difference between each polymorph within a specific polymorph type and its corresponding reference structure ($\Delta H_{\text{DFT}}$), while y-axis indicates Helmholtz vibrational free energy difference between them ($\Delta A_{\text{DFT}}$ or $\Delta A_{\text{MACE}}$). The enthalpies were obtained from DFT calculations at 0 K, while Helmholtz free energies were calculated from both DFT and the trained MACE model at 300 K and 1000 K, respectively. The black circles represent the scatter point ($\Delta H_{\text{DFT}}$, $\Delta A_{\text{DFT}}$), while the red triangles denote the scatter point ($\Delta H_{\text{DFT}}$, $\Delta A_{\text{MACE}}$). The grey region beneath the black dashed line ($y=-x$) indicates instances where the sum of $\Delta H_{\text{DFT}}$ and $\Delta A_{\text{DFT}}$ (or $\Delta A_{\text{MACE}}$) is negative, suggesting the polymorph within the region is more thermodynamically stable than the reference structure. Considering the potential errors in DFT calculations, a 10 meV offset is applied, expanding the possible region for potential transition to include both the blue region and the grey region below the dashed line ($y=-x+10$). If both DFT and MACE predict a polymorphic transition for specific materials (i.e. both data points lie within the shadowed region and they have the lowest Helmholtz free energy), they are marked with filled black circles and filled red triangles. Scatter plots (a) and (b) represent predictions for polymorphic transition occurrence at (a) 300 K and (b) 1000 K, respectively. Plots (c) and (d) provide zoomed-in views of (a) and (b), respectively.
	}
	\label{fig:phase-transition}
\end{figure*}

While the dynamical stability of a crystal structure indicates that the structure remains stable under small perturbations, it does not guarantee its thermodynamic stability. A crystal structure can be dynamically stable but thermodynamically unstable if there exists another structure with lower free energy under certain conditions, such as at higher temperatures. When external kinetic energy is applied to such a structure and the increased kinetic energy is enough to overcome energy barriers between different structural configurations, a phase transition may occur. Despite the importance of thermodynamic stability in the material discovery field, systematic computational studies exploring the stability of polymorphs are lacking in the literature. Using our phonon dataset containing a number of polymorphs, we conduct thermodynamic stability analysis and evaluate the model performance in predicting polymorphic transitions. The thermodynamic stability of a structure is determined by free energy, such as Gibbs free energy and Helmholtz free energy. In this study, we use Helmholtz free energy for thermodynamic stability analysis, which is defined under constant volume conditions, disregarding thermal expansion effects. The fixed volume of materials is obtained from DFT structure relaxation calculations performed at 0~K. Helmholtz free energy $A$ is expressed as follows at a given temperature $T$ and volume $V$ \cite{korotaev2018reproducibility}:
\begin{equation} \label{eq10}
A(V,T)=U-TS = U_0 + U_\text{vib} - TS_\text{vib} = U_0 + A_\text{vib}
\end{equation}
where $U$ is the internal energy of a system and $U_0$ specifically refers to the internal energy at a static state, corresponding to the system's energy when it is at 0~K. $U_\text{vib}$ represents a vibrational contribution induced by thermal lattice vibrations, and it includes the zero-point energy. $S_\text{vib}$ is the vibrational entropy of a system, and $A_\text{vib}$ is Helmholtz vibrational free energy. In this research, we neglect other entropic factors such as electronic and magnetic entropy, and focus on the vibrational entropy contributions on Helmholtz free energy under the harmonic approximation. 

To calculate the Helmholtz free energy $A$ of materials, we need the internal energy at a static state $U_0$ and the Helmholtz vibrational free energy $A_\text{vib}$. Helmholtz vibrational free energy can be calculated using Eq.\ref{eq6}. The internal energy $U$ is defined as $U = H - PV$, where $H$ is the enthalpy and $P$ is the pressure. In solids, the $PV$ term is significantly smaller than $H$ and can be disregarded. Since DFT calculations are performed at 0 K, we can approximate the $U_0$ as the enthalpy computed from DFT calculations, $H_\text{DFT}$. Consequently, Helmholtz free energy can be approximated as:
\begin{equation} \label{eq11}
A(V,T) \approx H_\text{DFT} + A_\text{vib}
\end{equation}
The Helmholtz vibrational free energy $A_\text{vib}$ is obtained from both MACE and DFT. For simplicity, we denote the Helmholtz vibrational free energies obtained from MACE and DFT calculations as $A_\text{MACE}$ and $A_\text{DFT}$, respectively. Although the MACE model also can be used for total potential energy predictions when it is trained with energies, in this study, we focus on the harmonic phonon property predictions. 

We computed two sets of $A_\text{vib}$ ($A_\text{MACE}$ and $A_\text{DFT}$) using phonon frequencies at 300 K and 1000K. As shown in Fig.\ref{fig:evaluation} (c) and (d), we compared $A_\text{MACE}$ with $A_\text{DFT}$ for each temperature case. The MAE of $A_\text{vib}$ at 300 K is 2.19 meV/atom with the $R^2$ values of 0.995, indicating a strong agreement between MACE predictions and DFT calculations. It is worth mentioning that such an MAE is much smaller than previous models \cite{gurunathan2023rapid, legrain2017chemical}. At 1000K, the MAE increases to 9.30 meV/atom. Although this error is relatively larger, the $R^2$ value is 0.982, demonstrating overall good agreement. With increasing temperature, the absolute values of $A_\text{vib}$ also increase, leading to the increased absolute error. 

We then assessed model performance in predicting thermodynamic stability for polymorphs in the phonon dataset. Polymorphs are different crystal structures that exist at a given chemical composition under different external conditions. In our phonon dataset, there are a total of 126 polymorphs distributed across 49 distinct polymorph types. The 49 types of polymorphs and the number of polymorphs of each specific type are described in the Fig.\ref{fig:polymorph} (a). Thermodynamic stability was determined by computing the difference in Helmholtz free energy ($\Delta A \approx \Delta H_\text{DFT} + \Delta A_\text{vib}$). For each polymorph type, a reference structure is chosen based on the enthalpy $H_\text{DFT}$, where the reference structure has the lowest enthalpy and is considered as a stable structure without considering the vibrational contribution. For example, within the NiAs polymorph type, the reference structure is ``NiAs-66120", because it has the lowest enthalpy compared to other polymorphs ``NiAs-29303" and ``NiAs-611024". The number, such as 66120, denotes the index of the Inorganic Crystal Structure Database (ICSD). To evaluate the model's performance in predicting the thermodynamic stability of polymorphs, we calculated both $\Delta A_\text{DFT}$ and $\Delta A_\text{MACE}$ as the difference between the Helmholtz vibrational free energy of the polymorph and that of the reference structure, i.e., $A_\text{vib} - A_{\text{vib}_\text{ref}}$. As shown in Fig.\ref{fig:polymorph} (b) and (c), we compared Helmholtz vibrational free energy difference, $\Delta A_\text{DFT}$ versus $\Delta A_\text{MACE}$, at 300 K and 1000 K. The MAE of $\Delta A_\text{MACE}$ compared with $\Delta A_\text{DFT}$ at 300 K is 4.27 meV/atom with the 0.778 of $R^2$ and it indicates that the model predictions are in overall good agreement with DFT. However, as shown in Fig.\ref{fig:polymorph} (c), the MAE increases to 20.83 meV/atom at 1000 K and the $R_2$ decreases to 0.315. Although lots of data points align with the $y=x$ guidelines, some of them show large variation, especially, when the absolute value of $\Delta A_\text{vib}$ increases. Despite the increased error in predictions at higher temperatures, the MAE of 20.83 meV/atom is still much smaller than the quantum chemical error of 43 meV/atom \cite{bogojeski2020quantum}. 

For the thermodynamic stability analysis, we also need to calculate the $\Delta H_\text{DFT}$. Similarly to the $\Delta A_\text{vib}$ calculations, $\Delta H_\text{DFT}$ was calculated based on the enthalpy of a reference structure for each polymorph type ($\Delta H_\text{DFT} = H_\text{DFT} - H_{\text{DFT}_\text{ref}}$). Note that here we only have $\Delta H_\text{DFT}$ obtained from DFT calculations, while $\Delta A_\text{vib}$ was obtained from both DFT and MACE. In the Fig.\ref{fig:phase-transition} (a)-(d), red triangles represent data points ($\Delta H_\text{DFT}$, $\Delta A_\text{MACE}$), while black circles indicate data points ($\Delta H_\text{DFT}$, $\Delta A_\text{DFT}$). Hence, each red triangle and its corresponding black circle share the same x-value, which is $\Delta H_\text{DFT}$. If $\Delta A_\text{MACE}$ and $\Delta A_\text{DFT}$ values are very close, the scatter points will overlap. Conversely, if there is a significant difference between their values, the scatter plots of red triangles and black circles will be far apart along the y-axis ($\Delta A_\text{vib}$). Fig.\ref{fig:phase-transition} (a) and (b) show the potential occurrence of polymorphic transition at 300K and 1000 K, respectively, and (c) and (d) offer an enlarged view of (a) and (b), correspondingly. The grey dashed lines in the figures represent the guide for $y + x = 0$. The shaded region below the lines corresponds to $\Delta A \approx \Delta H_\text{DFT} + \Delta A_\text{vib} < 0$, indicating that the polymorph is more thermodynamically stable than the reference structure at the given temperature. However, there is a possibility of polymorphic transition occurrence even if $\Delta A$ is positive, due to the theoretical limitations of DFT compared to experimental measurements. To account for potential errors in DFT, we apply an offset of 10 meV. If $\Delta A$ of a polymorph is less than 10 meV, meaning the data point lies within the shaded region including both the blue and grey areas in Fig.\ref{fig:phase-transition}, we classify the polymorph as more thermodynamically stable than the reference structure within the same polymorph type. Additionally, it is necessary to compare the polymorphs against structures other than the reference to identify the polymorphs with the lowest Helmholtz free energy. If both polymorphs A and B are more thermodynamically stable than the reference C, we compare the $\Delta A$ values of A and B. If the $\Delta A$ of polymorph A is smaller than that of polymorph B, then polymorph A is thermodynaically stable at the given temperature, and the polymorphic transition may occur. In Fig.\ref{fig:phase-transition} (c) and (d), polymorphs with both DFT and MACE data points within the shaded region and have the lowest $\Delta A$ in each polymorph type are represented with filled markers. This indicates that both DFT and MACE predict a potential for polymorphic transitions. 

At a temperature of 300 K, both DFT and MACE predict the possibility of 19 potential polymorphic transitions among the polymorphs. Remarkably, 16 of these potential transitions are consistent between the predictions made by DFT and MACE. On the other hand, at 1000 K, DFT and MACE predict 19 and 18 potential polymorphic transitions, respectively. Among these, 13 potential transitions are common to both methods. The lists of materials that consistently exhibit the potential for polymorphic transitions according to both DFT and the MACE model are provided in Appendix \ref{sec:appendix}. These findings represent a high degree of agreement between DFT and MACE regarding the likelihood of polymorphic transitions at both 300 K and 1000 K.
Our trained MACE model shows good performance in predicting polymorphic transitions, even though the MAE of $\Delta A_\text{DFT}$ is 4.27 meV/atom and 20.83 meV/atom at temperatures of 300 K and 1000 K.
\section{\label{sec:conclusion} Conclusion}

We have proposed an approach to accelerate high-throughput phonon calculations using machine learning universal potentials and an efficiently constructed training dataset. We used a state-of-the-art MACE machine learning model and trained it using our comprehensive training dataset. To generate the training dataset, we created approximately six supercells for each of the 2,738 unary or binary materials covering a total of 77 elements across the periodic table. We randomly perturbed all atoms within the supercells with atomic displacements ranging from 0.01 to 0.05~\r{A} to collect extensive force information. In total, 15,670 crystal structures and 8.1 million force components of the structures were used as the training dataset. During the training process, we only used forces without energies to train the MACE model to achieve highly accurate force predictions. 

As a result, our trained MACE model achieves an impressive force prediction MAE of 18.8 meV/\r{A} on the training dataset and 20.5 meV/\r{A} on the validation dataset, respectively, significantly outperforming previous models. The model also demonstrates excellent performance in predicting phonon frequencies on the phonon dataset consisting of 384 materials, with an MAE of 0.18 THz compared to DFT calculations. It accurately predicts the dynamical stability of materials with an 86.2\% accuracy, suggesting its potential as a powerful tool for the initial screening of dynamically unstable structures. We also evaluated the performance of the trained MACE model in predicting the thermodynamic stability. It shows good agreement with DFT calculations for Helmholtz vibrational free energy with MAEs of 2.19 meV/atom and 9.30 meV/atom at 300 K and 1000 K, respectively. Additionally, the model shows promising capability in predicting potential polymorphic transitions at both 300 K and 1000 K. 

Our findings highlight the potential of MLIPs, specifically the MACE model, for accurately predicting harmonic phonon properties when trained with an efficiently built training dataset. By leveraging the machine learning approach, we can significantly reduce computational costs associated with constructing a training dataset. When constructing our phonon dataset with 384 materials, we employed Phonopy package to automatically generate supercells with finite displacements. Phonopy considers the symmetry of structures to determine the minimum number of required supercells for phonon calculations. As a result, the number of required supercells ranged from 1 to 132 depending on the structures, with an average of 13.3. Remarkably, 47\% of materials in the dataset required more than six supercells, with low-symmetry materials demanding over 100 supercells. However, by constructing a training dataset using six supercells for each material and leveraging the MACE model, we achieved accurate predictions of harmonic phonon properties across a wide range of materials, thus demonstrating the efficiency of our approach. Moreover, if computational resources allow, the model's accuracy is expected to improve systematically as more training data becomes available.

In addition to our primary focus on accelerating harmonic phonon calculations, we also provide a comprehensive dataset comprising 2,738 unary and binary materials across 77 elements, computed using DFT. This extensive dataset contains energies as well as forces for 15,670 crystal structures generated from 2,738 materials, providing valuable information for a wide range of computational studies. Moreover, the dataset includes high-quality phonon data for 384 structures. It is important to note that our dataset is not limited to the MACE machine learning model alone. Researchers can leverage this dataset to enhance other machine learning interatomic potential models to predict total energies, forces, and especially vibrational properties.

While we restrict the phonon predictions under harmonic approximations in this study, anharmonic effects are also crucial for understanding the thermal properties of materials. Future work should focus on incorporating anharmonic effects into the machine learning framework to improve the accuracy of phonon predictions, particularly in systems where anharmonic contributions play a significant role in. Another limitation of our study is our training dataset only covers unary or binary materials. The trained MACE model may not perform as well in predicting phonon properties of ternary, quaternary, or other complex materials. Expanding the training dataset to include more diverse structures could enhance the model's applicability and accuracy across a broader range of materials.

\textbf{Acknowledgements.}
H. L. and Y. X. acknowledge support from the US National Science Foundation through award 2317008. We acknowledge the computing resources provided by Bridges2 at Pittsburgh Supercomputing Center (PSC) through allocations mat220006p and mat220008p from the Advanced Cyber-infrastructure Coordination Ecosystem: Services \& Support (ACCESS) program, which is supported by National Science Foundation grants \#2138259, \#2138286, \#2138307, \#2137603, and \#2138296.

\textbf{Data Availability Statement}
The data that support the findings of this study, including datasets, machine learning models, and python scripts, will be published upon the acceptance of the manuscript.


\section{\label{sec:appendix} Appendix}

\appendix
\subsection{MACE hyperparameters}

\begin{table}[b!]
    \centering
    \begin{tabular}{c|c|c|c}
    \hline
    $S$ & MAE E (meV/atom) & MAE F (meV/\r{A}) & GPU (hrs)\\
    \hline
    1 & 24.6 & 28.4 & 13.6 \\ 
    2 & 25.7 & 27.3 & 23.7 \\ 
    3 & 21.0 & 25.9 & 32.5 \\
    \hline
    \end{tabular}
    \caption{Effect of number of interaction layers on MAE of energy and force and computational cost of GPU}
    \label{tab:nlayer}
\end{table}

\begin{table}[b!]
    \centering
    \begin{tabular}{c|c|c}
    \hline
    Learning rate & MAE E (meV/atom) & MAE F (meV/\r{A})\\
    \hline
    0.01 & 28.8 & 37.0 \\ 
    0.02 & 31.1 & 37.5 \\ 
    0.05 & 45.4 & 51.0 \\
    \hline
    \end{tabular}
    \caption{Effect of learning rate on MAE of energy and force}
    \label{tab:learning-rate}
\end{table}

\begin{table}[b!]
    \centering
    \begin{tabular}{c|c|c}
    \hline 
    Loss weight(F:E) & MAE E (meV/atom) & MAE F (meV/\r{A})\\
    \hline
    50 : 10 & 17.2 & 27.6 \\
    100 : 10 & 25.7 & 27.3 \\ 
    1000 : 10 & 51.2 & 25.1 \\ 
    1000 : 0 & 400.6 & 23.4 \\ 
    \hline
    \end{tabular}
    \caption{Effect of force and energy weights in the loss function on MAE of enegy and force}
    \label{tab:weight-ratio}
\end{table}

Optimal hyperparameter settings for a GNN can enhance training efficacy and prediction accuracy \cite{yuan2021novel}. In our efforts to obtain optimal hyperparameters for training the MACE machine learning model, we conducted a series of tests involving the number of layers ($S$), learning rate, and weights of energy and force in the loss function, using our training dataset. We maintained consistency by fixing other hyperparameters: $r_{\text{cut}}$ = 6\r{A}, $\nu$ = 3, $L_\mathrm{max}$ = 0, and $k$ = 64. 

Our tests on the number of interaction layers were conducted using a default learning rate of 0.01, with force and energy weights set to 100 and 10, respectively. Table. \ref{tab:nlayer} presents the MAE of energy and force after 150 epochs. It shows increasing the number of interaction layers correlates with a reduction in the MAE of force. However, this improvement is quite small, considering the significant increase in GPU hours required. Table \ref{tab:learning-rate} illustrates the impact of different learning rates on the MAE of energy and force after training 100 epochs. Two interaction layers were used, and the force and energy weights in the loss function were set to 100 and 10. While the default learning rate for MACE is 0.01, adjusting the learning rate can sometimes improve training effectiveness. However, our results indicate that the learning rates of 0.02 and 0.05, resulted in larger MAE values for our dataset. Similarly, Table \ref{tab:weight-ratio} examines the effect of weights of energy and force in the loss function on the MAE of energy and force after 150 epochs, with two interaction layers and a default learning rate. Notably, increasing energy weight significantly reduce the MAE of energy, however it leads to slight increase of the MAE force.

Consequently, considering the effectiveness of training and accuracy, we opted for two interaction layers for the final MACE model training. A learning rate of 0.01 was used, and force weight was exclusively applied to the loss calculations to achieve a high level of accuracy for force predictions.

\subsection{Performance in heat capacity and entropy predictions}

\begin{figure*}[htpb]
	\includegraphics[width = 0.85\linewidth]{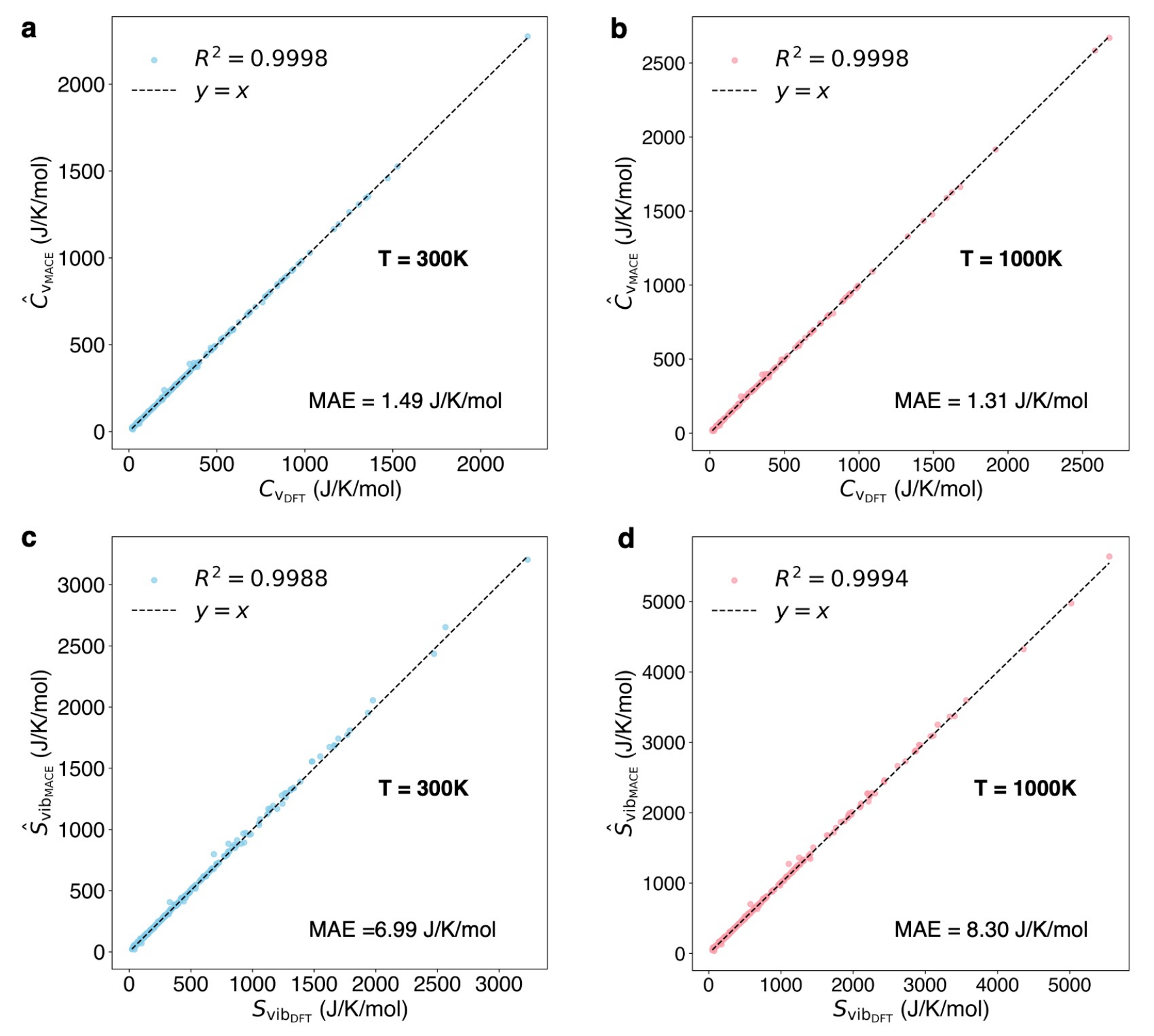}
	\caption{Comparison of the trained MACE model's performance in predicting constant volume heat capacity and vibrational entropy using the phonon dataset. (a) and (b) show the comparison of predicted constant volume heat capacity by our trained MACE model with DFT-calculated values at 300 K and 1000 K, respectively. Similarly, (c) and (d) illustrate the comparison of vibrational entropy between the MACE model and DFT at 300 K and 1000 K, respectively.
	}
	\label{fig:thermal-eval}
\end{figure*}

Similar to the calculation of Helmholtz vibrational free energy, we also can compute the constant volume heat capacity ($C_v$) and vibrational entropy ($S_\text{vib}$) under harmonic approximation as follows \cite{Dove1993, Phonopy_first}:

\begin{equation} \label{eq7}
C_v = \sum_{\textbf{q} \nu} k_B \left(\frac{\hbar \omega_{\textbf{q} \nu}}{k_{B} T}\right)^2 \frac{\exp(\hbar \omega_{\textbf{q} \nu}/k_BT)}{[\exp(\hbar \omega_{\textbf{q} \nu}/k_BT) - 1]^2}
\end{equation}

\begin{equation} \label{eq12}
\begin{split}
S_\text{vib} & = \frac{1}{2T} \sum_{\textbf{q} \nu} \hbar \omega_{\textbf{q} \nu}\text{coth}(\hbar \omega_{\textbf{q} \nu} / 2 k_{B} T) \\
& - k_{B} \sum_{\textbf{q} \nu} \ln [2 \text{sinh} (\hbar \omega_{\textbf{q} \nu} / 2 k_{B} T)]
\end{split}
\end{equation}

The trained MACE model demonstrates high accuracy in predicting heat capacity at constant volume and vibrational entropy. At 300 K, the MAE for heat capacity at constant volume is 1.49 J/(mol$\cdot$K), while at 1000 K, it is 1.31 J/(mol$\cdot$K), as shown in Fig. \ref{fig:thermal-eval} (a) and (b). In both cases, the $R_2$ values are close to 1.0, indicating excellent agreement between the MACE model predictions and calculated values via DFT simulations. Similarly, fig. \ref{fig:thermal-eval} (c) and (d) present the case for vibrational entropy, and the MAE is 6.99 J/(mol$\cdot$K) at 300K and 8.30 J/(mol$\cdot$K) at 1000 K, with $R_2$ values approaching 1.0. These results are comparable to the reported values of 1.58 J/(mol$\cdot$K) for heat capacity at constant volume and 7.26 J/(mol$\cdot$K) for vibrational entropy at room temperature reported by the ALIGNN model \cite{gurunathan2023rapid}. These findings suggest the accuracy and reliability of the MACE model in capturing the thermodynamic properties of the system.

\subsection{Polymorphic transitions}

We utilized both DFT and the trained MACE model to investigate the occurrence of polymorphic transitions within the phonon dataset materials at temperatures of 300 K and 1000 K. We provide lists of materials that are consistently identified as potentially thermodynamically stable by both DFT and the MACE model. This consistency demonstrates the accuracy of the MACE model predictions for polymorphic transitions. However, it is important to note that we applied an offset of 10 meV to allow for DFT accuracy and also conducted our polymorphs analysis within the phonon dataset. Certain polymorphs, potentially more stable than those listed, may not be included in the phonon dataset.

Materials identified as potentially stable by both DFT and MACE at 300 K (total 16) are:
`NiAs-29303', `AlAu$_2$-57497', `ZrSi-16771', `WS$_2$-202367', `SmTl-106062', `Li$_3$Sb-26879', `Ce-41824', `NaN$_3$-34676', `TiSi$_2$-168419', `CuS$_2$-53328', `GaN-41546', `HgI$_2$-281133', `SrH$_2$-69077', `SbI$_3$-26082', `VB$_2$-165125', `Al$_5$Mo-105520'.

Materials identified as potentially stable by both DFT and MACE at 1000 K (total 13) are:
`NiAs-29303', `AlAu$_2$-57497', `ZrSi-16771', `WS$_2$-202367', `Li$_3$Sb-26879', `Ce-41824', `NaN$_3$-34676', `TiSi$_2$-168419', `GaN-41546', `SrH$_2$-69077', `SbI$_3$-26082', `VB$_2$-165125', `Al$_5$Mo-105520'.

\bibliography{MLHP}

\end{document}



\widetext
\begin{center}
	\textbf{\large Supplementary Material: Accelerating High-Throughput Phonon Calculations Using Machine Learning-Based Universal Potentials}
\end{center}

\begin{center}

Huiju Lee$^{1}$, Vinay I. Hegde$^{2}$, Chris Wolverton$^{2}$, and Yi Xia$^{1}$

\vspace{0.3cm}

\text{$^1$ Department of Mechanical and Materials Engineering,}
\text{Portland State University,Portland, OR 97201, USA}

\text{$^2$ Department of Materials Science and Engineering, Northwestern University, Evanston, IL 60208, USA}

\end{center}


\setcounter{equation}{0}
\setcounter{figure}{0}
\setcounter{table}{0}
\setcounter{page}{1}
\makeatletter
\renewcommand{\thepage}{S\arabic{page}}
\renewcommand{\theequation}{S\arabic{equation}}
\renewcommand{\thefigure}{S\arabic{figure}}
\renewcommand{\thetable}{S\arabic{table}}
\renewcommand{\bibnumfmt}[1]{[S#1]}
\renewcommand{\citenumfont}[1]{S#1}



\section{Comparing phonon dispersions: Trained MACE machine learning model vs. DFT calculations}

Full phonon dispersions were computed for 384 materials in our phonon dataset to assess the performance of the trained MACE machine learning model compared to DFT calculations. In the following plots, black solid lines indicate the DFT results, while blue dashed lines represent phonon dispersions computed from the trained MACE machine learning model. The chemical formula and the index of the Inorganic Crystal Structure Database (ICSD) for each material are displayed on each phonon dispersion plot.

\foreach \i in {1, 2,3,...,10}{
    \begin{figure*}[htp]
        \includegraphics[width=0.94\linewidth]{dispersion_\i.pdf}
        \label{fig:dispersion_\i}
    \end{figure*}
}


\bibliography{MLHP}